\documentclass[proof]{WileyASNA-v1}

\articletype{Review Article}%

\received{19 September 2018}
\revised{11 March 2019}
\accepted{15 March 2019}

\begin{document}

\title{How stars and planets interact: a look through the high-energy window}

\author{Katja Poppenhaeger$^{1,\,2,\,3}$}

\address[1]{\orgname{Leibniz Institute for Astrophysics Potsdam}, \orgaddress{An der Sternwarte 16, 14482 Potsdam, \country{Germany}}}

\address[2]{\orgdiv{University of Potsdam}, \orgname{Institute of Physics and Astronomy}, \orgaddress{Karl-Liebknecht-Stra{\ss}e~24/25, 14476 Potsdam, \country{Germany}}}

\address[3]{\orgdiv{Queen's University Belfast}, \orgname{Astrophysics Research Centre}, \orgaddress{BT7 1NN Belfast, \country{United Kingdom}}}

\corres{*\email{kpoppenhaeger@aip.de}}


\abstract{The architecture of exoplanetary systems is often different from the solar system, with some exoplanets being in close orbits around their host stars and having orbital periods of only a few days. In analogy to interactions between stars in close binary systems, one may expect interactions between the star and the exoplanet as well. From theoretical considerations, effects on the host star through tidal and magnetic interaction with the exoplanet are possible; for the exoplanet, some interesting implications are the evaporation of the planetary atmosphere and potential effects on the planetary magnetism. In this review, several possible interaction pathways and their observational prospects and existing evidence are discussed. A particular emphasis is put on observational opportunities for these kinds of effects in the high-energy regime.}


\fundingInfo{}

\maketitle

\section{Introduction}\label{sec1}

\begin{figure}
\includegraphics[width=0.49\textwidth]{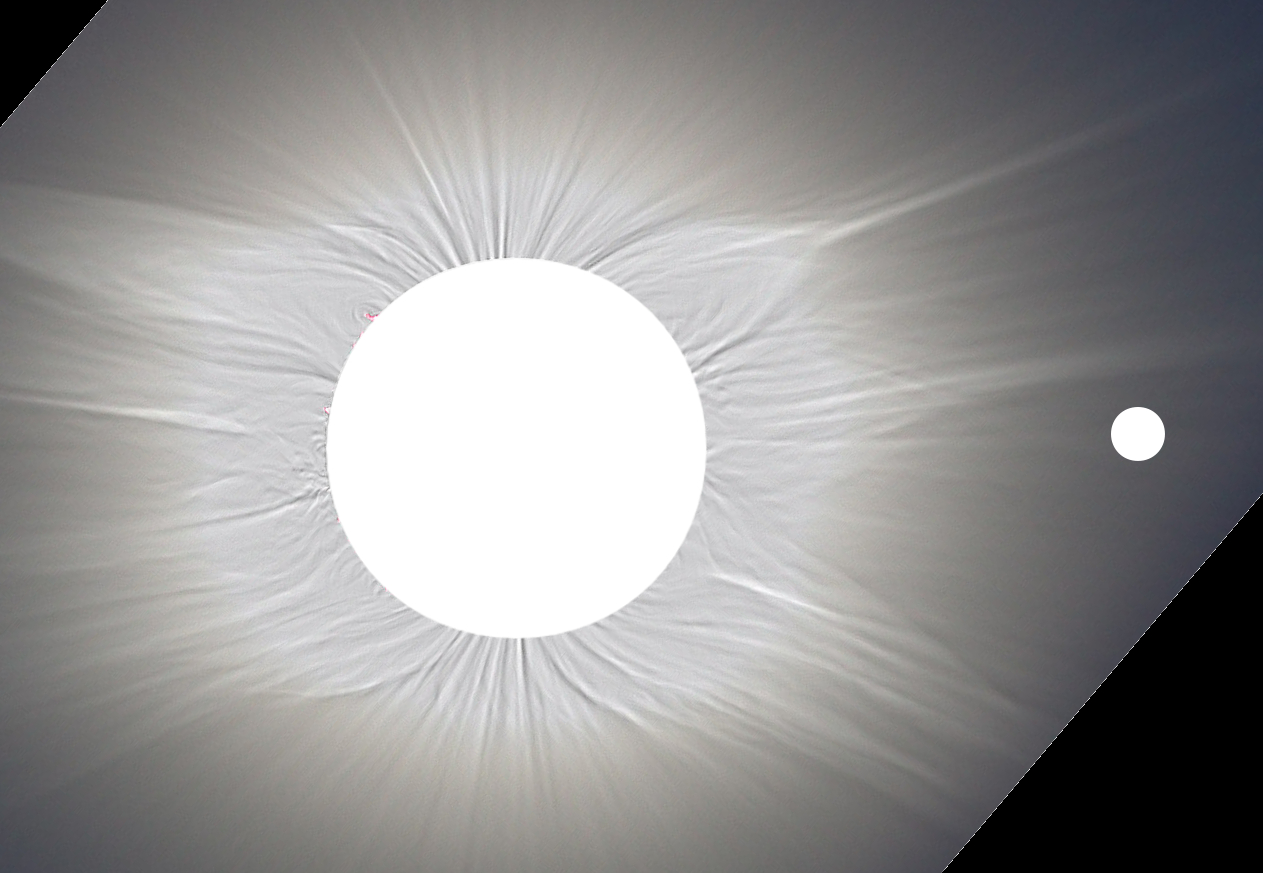}
\caption{Montage of a close-in Hot Jupiter orbiting a cool star with the solar corona as background, all object sizes and distances to scale (taking the values of the WASP-12 exoplanet system as an example). Some exo\-pla\-nets are close enough to their host stars to orbit within the outer structures of the stellar corona. Solar corona image by K.\ van Gorm, montage by K.\ Poppenhaeger.}
\label{fig1}
\end{figure}

Exoplanets can be found in orbits around their host stars that are not observed in our own solar system. The first exoplanet discovered around a solar-like star, 51~Peg~b, has an orbital period of less than five days \citep{MayorQueloz1995}, and planets of this type were subsequently named "Hot Jupiter". Other exoplanet systems such as the WASP-12 system \citep{Hebb2009} have a large planet orbiting the host star within roughly one day, corresponding to a semi-major axis of only a few stellar radii (see Fig.~\ref{fig1}). Interactions between such an extremely near-by planet and the host star may therefore be expected to occur; both the planet and the star may experience observable effects from different kinds of interaction. Giving particular attention to observable effects in the high-energy regime, this review will first discuss effects on the host star and then effects on the planet.

\section{Planetary effects on host stars}\label{sec2}

Most exoplanets are detected around cool stars on or near the main sequence. This is typically because the main detection methods, radial velocity and transits, work better around small stars. This leads to a known exoplanet zoo in which most exoplanet host stars have an outer convective envelope or are fully convective. Amongst these stars a range of magnetic phenomena are very common and are collectively called ``magnetic activity'' (for a recent review see \citealt{Testa2015}). This includes the existence of a chromosphere and corona, starspots, stellar winds, stellar flares, and coronal mass ejections. Ultimately these phenomena are driven by a stellar dynamo, which uses the convection of the stellar material in the outer layers of the star to convert stellar rotation into highly localized and time-variable magnetic field structures (for a review see \citealt{Charbonneau2014review}). These are observable in great detail for the Sun, where for example coronal loops can be observed to change over time. For stars other than the Sun, where the stellar surface is not spatially resolved in observations, the coronal emission and the time evolution of flares is observable at X-ray wavelengths, for example.

The possible effects near-by exoplanets might have on such cool stars divide into two main pathways, tidal and magnetic interaction \citep{Cuntz2000}, which are outlined below.

\subsection{Tidal interaction in star-planet systems}

Stars and giant planets influence each other through tides, i.e.\ the star raises tides on its gaseous planet and vice versa. 
When the planetary orbit and the stellar spin do not have the same period, then tides raised by the planet on the star can have an influence on the star itself. Typically, stars (meaning cool stars in the context of exoplanet host stars) start out with a short rotation period of a few days when they arrive on the main sequence, and continually spin down through the process of magnetic braking. Many detected Hot Jupiters, with their orbital periods having a duration of up to a few days, orbit older stars that have a longer rotation period than the planetary orbital period. In such systems the tidal lag raised by the planet on the star lags behind the planet on its orbit, and the gravitational pull of the planet on the bulge transfers angular momentum from the planetary orbit into the stellar spin. In addition to this interaction through the so-called equilibrium tide \citep{Zahn2008, Remus2012}, there is an additional part to the interaction called the dynamical tide. This concerns the excitation of eigenmodes of the star due to the gravitational potential of the planet together with the Coriolis force of the star \citep{Ogilvie2007}, and depends sensitively on stellar properties such as the thickness of the outer stellar convection zone. Both tides together govern how the exoplanetary orbit evolves over time, and how the stellar rotation is influenced by it. While equilibrium tides have been investigated for a while in this context \citep{Ferraz-Mello2008, Levrard2009, Jackson2016}, there is recent progress in adding dynamical tides to the picture for cool exoplanet host stars \citep{Mathis2015, Bolmont2017, Heller2018}.

Observationally, studies  of  several individual star-planet systems  have  shown  that  some  Hot  Jupiter  host  stars  are  rotating
faster than expected \citep{Pont2009, Brown2011}.  In order to control for stellar age (which is typically hard to measure for field stars) and therefore determine the expected unperturbed rotational state of the star, \citet{Poppenhaeger2014} investigated wide pairs of common proper motion stars where one star hosts an exoplanet, to determine if these presumably same-age stars show deviations in their rotational state and magnetic activity; star-planet systems with expected strong tidal interaction between star and planet showed elevated X-ray activity levels compared to the second star in the system. 

It is important to point out that the full population of known exoplanets can display trends of stellar activity with planetary parameters that are not necessarily caused by star-planet interaction. Since exoplanets are more easily detected around stars with low activity, some detection biases are present in the sample \citep{Kashyap2008, Poppenhaeger2010, Poppenhaeger2011}; only for very close Hot Jupiters do the trends seem to be caused by actual physical interactions instead of sampling biases \citep{Miller2012}. However, the picture still remains unclear and requires more investigation; for example, there are suggestions that Hot Jupiters may even suppress the magnetic dynamo for a star with a very thin outer convective layer, such as WASP-18 \citep{Pillitteri2014WASP18}.

\subsection{Magnetic interaction in star-planet systems}

Another intriguing possibility is the interaction of the stellar and the planetary magnetic field. A template for a possible interaction of this kind is Jupiter and its moon Io in the solar system, which act as a unipolar inductor and cause X-ray and ultraviolet (UV) emitting aurorae near Jupiter's poles \citep{Goldreich1969}. It has been hypothetized that Hot Jupiters might cause hot spots in the stellar chromosphere and corona through an interaction of their magnetospheres, for example through reconnection of the stellar and planetary magnetic field lines. The first observational search for such effects was conducted by \citet{Shkolnik2005}, who found that out of a sample of 13 stars with Hot Jupiters two displayed a modulation of the stellar chromospheric Ca\,II H and K line flux that was modulated with the planetary orbital period and not the stellar rotation period. However, subsequent follow-up of targets showed that only modulation with the stellar rotation period could be observed at later epochs \citep{Shkolnik2008}; see also \citet{PoppenhaegerLenz2011}. Since stellar activity is an intrinsically time-variable phenomenon even without planets, detecting such effects in an unambiguous manner is difficult. Theoretical considerations also showed that hot spots might not necessarily be found at the sub-planetary point, complicating things further \citep{Lanza2008}. Other repeated observations of stellar X-ray and far-UV fluxes showed increases after the planetary eclipse (not transit), which have been interpreted as coronal flux enhancements due to planetary material falling onto the star and being evaporated into the corona, increasing the coronal emission measure temporarily \citep{Pillitteri2014, Pillitteri2015}. It is therefore currently unclear at which point of the planetary orbit one should actually expect to see a flux increase from potential planet-induced hot spots.

A possible way out of the observational conundrum is to reduce the ambiguity in expectations when an X-ray flux increase should occur. Systems which allow us to form time-constrained expectations are ones that host planets in eccentric orbits. For binary stars in eccentric orbits, X-ray flaring at the time of periastron passage has been observed \citep{Getman2011}; however a caveat should be noted, namely that in a larger sample the statistical evidence is still somewhat tentative \citep{Getman2016}, and an absence of flaring at periastron does not necessarily imply absence of magnetic interaction. In the case of eccentric star-planet systems, the HD~17156 system has received detailed observational attention in the soft X-ray band with \textit{XMM-Newton}. The system hosts a massive planet in a highly eccentric ($e=0.68$) orbit of roughly 20 days orbital period. The system was observed twice during periastron passage of the planet and twice during apoastron, and in all cases the system was only successfully detected in X-rays during the periastron passage \citep{Maggio2015}. While more coverage of any such system would be desirable to strengthen an interpreatation of planetary influence over stochastic stellar variability, this is a strong indication of observable star-planet interaction.

\section{Stellar effects on planets}\label{sec3}

While planetary effects on host stars present some difficulties in observational confirmation due to the reasons outlined above, the effect of stars on planets in their vicinity has been measured successfully in a number of cases. The effect of stellar irradiation on exoplanetary atmospheres and the potential implications for planetary magnetic fields are discussed here.

\subsection{Planetary evaporation}

Stellar irradiation of a planet determines the equilibrium temperature of the planet. While the bolometric stellar irradiation (together with greenhouse effects due to a planetary atmosphere) influences the surface temperature of the planet, irradiation with high-energy photons from the star can have a more immediate effect in the planetary atmosphere itself. \citet{Yelle2004} first described comprehensively how atmospheres of planets fare under high-intensity stellar irradiation. High-altitude layers of the atmosphere are strongly heated by extreme ultra-violet and X-ray photons, which can lead to mass-loss of the exoplanetary atmosphere into space. \citet{Murray-Clay2009} calculated that especially stars with strong magnetic activity can drive planetary mass loss rates of up to a few $10^{12}$\,g/s. Such a mass loss rate typically does not cause a large dent in the atmosphere budget of a Hot Jupiter, because the stellar high-energy emission decreases exponentially over time, and the integration over the lifetime of the system quickly causes a saturation of the total atmospheric mass lost; see for example the expected mass loss history of the Hot Jupiter HD~209458~b's \citep{Penz2008PSS} or the very low current X-ray brightness of the Hot Jupiter host star 51~Peg \citep{Poppenhaeger2009}. In contrast, smaller planets with thinner atmospheres might lose significant portions of their atmosphere or even lose it completely \citep{Sanz-Forcada2010, Sanz-Forcada2011, Lopez2012, Poppenhaeger2012, Lalitha2014}. 

Some direct observational evidence for ongoing evaporation of exoplanet atmospheres has been presented for several planets over the past years. Such observations make use of ultra-violet observations during exoplanetary transits in order to detect escaping hydrogen from the planet as an absorption feature in the hydrogen Ly-$\alpha$ line emitted by the host star. Ongoing mass loss has been reported for example for the Hot Jupiters HD~209458~b \citep{Vidal-Madjar2003}, HD~189733~b \citep{Lecavelier2010}, and a spectacularly large absorption feaure, causing an occultation of almost half the stellar flux in the wings of the Ly-$\alpha$ line was discovered by \citet{Kulow2014} for the Warm Neptune GJ~436~b and re-investigated in more detail by \citep{Ehrenreich2015}. Additionally, observational evidence for the existence of extended upper atmosphere layers has been derived from UV observations \citep{Fossati2010} and also from soft X-ray observations through transit measurements with \textit{XMM-Newton} and \textit{Chandra} \citep{Poppenhaeger2013}. 

A task to be solved by high-energy observations is to study the evolution of stellar activity over the stellar lifetime in more detail, as the stellar high-energy output is the main driver for atmospheric mass loss. Recent efforts have made progress in studying the rotation and activity of cool stars, especially M dwarfs \citep{Stelzer2013, Stelzer2016}; the age-rotation evolution of stars similar to the Sun \citep{Barnes2016, vanSaders2016}; and the evolution of X-ray emission versus stellar age determined by asteroseismic measurements \citep{Booth2017}.

\subsection{Planetary magnetism}

Several planets in the solar system produce magnetic fields. The Juno probe has recently measured Jupiter's magnetic field in detail \citep{Moore2018}, but other observables such as auroral emission at the planetary poles are observable in soft X-rays for some solar system planets (for a review see \citealt{Bhardwaj2007}). Systems in which an exoplanet orbits the host star at a close distance differ from the solar system due to the strong irradiation and high surface temperature of the planet. Quite surprisingly, \citet{Christensen2009} showed that objects that generate magnetic fields through a convection-driven dynamo, such as planets, rapidly rotating low-mass stars, and contracting young stars, all fit the same scaling law in which the energy flux available for generating the magnetic field determines the field strength. While the field strengths for solar-system planets and for stars have been measured, Hot Jupiters have not yet led to a detection of their magnetic field, although radio signatures are expected from the interaction of the stellar wind with the planetary magnetic field \citep{Griessmeier2007} and some attempts have been made to detect those (see for example \citealt{Smith2009, Lecavelier2013}). However, from the aforementioned scaling law one may expect strong magnetic fields for very massive Hot Jupiters; it is important to note that one can expect relatively rapid rotation for planets in very close orbits since they will be rotationally locked to their orbital period, which can be shorter than a day for some exoplanets. Strong magnetic fields may therefore be expected, and \citet{Yadav2017Bfield} discuss the possibility of localized structures on the planetare surface of Hot Jupiters with fields strengths up to the kilogauss level. Also entirely different dynamo processes are possible for heated gas planets, such as atmospheric dynamos \citep{Rogers2017}. 

In parallel to X-ray emission from young brown dwarfs \citep{Stelzer2006BD, Berger2010}, also Hot Jupiters may be able to generate some form of high-energy emission associated with their magnetic dynamos, and next-generation high-energy observatories may reveal such processes through detection experiments. Particularly systems in which no X-ray emission is expected from the host star, such as Hot Jupiters around intermediate-mass stars (early to mid A-type stars, for example), will be suitable laboratories for such investigations.

\subsection{Outlook}

Current-generation X-ray telescopes like \textit{XMM-Newton} and \textit{Chandra} have played an important role in the exploration of interactions between stars and exoplanets, alongside UV observations with the Hubble Space Telescope \textit{HST} and ground-based optical observations at high spectral resolution. To study high-energy signatures of interactions between stars and planets, high flux sensitivity is important, as many of these signatures can barely be teased out of the current instrumentation at X-ray wavelengths. Upcoming or proposed high-energy missions such as \textit{Athena} \citep{Athena2013}, \textit{Arcus} \citep{Arcus2016} or \textit{LynX/X-ray Surveyor} \citep{Lynx2015} will provide new opportunities to study these exciting phenomena.


\section*{Acknowledgments}

K.P.\ acknowledges financial support from STFC through the Consolidated Grant scheme.

\bibliography{Poppenhaeger.bbl} 


\end{document}